\documentclass[prl,twocolumn,superscriptaddress,showpacs]{revtex4} 

\usepackage{graphicx}
\usepackage{dcolumn}
\usepackage{bm}
\usepackage{epstopdf}
\def\be{\begin{equation}}
\def\ee{\end{equation}}
\newcommand{\zp}    {\mbox{$Z ^{\prime}$}} 
\newcommand{\zpsm}    {\mbox{$Z ^{\prime}_{\rm SM}$}}

\newcommand{\mX}    {\mbox{$m_{X}$}}
\newcommand{\mll}    {\mbox{$m_{\ell\ell}$}} 

\newcommand{\ptm}    {\mbox{$p_{\rm{T}}$}}
\newcommand{\etm}    {\mbox{$E_{\rm{T}}$}}
\newcommand{\met}   {\mbox{$\not\!\! E_{T}$}}

\newcommand{\lims}   {\mbox{$\sigma(X_{\ell\ell})$ limit}}
\newcommand{\xlims}   {\mbox{$\sigma(X_{\ell\ell})$}}

\newcommand{\sigxll}   {\mbox{$\sigma(p\bar{p} \rightarrow X) \cdot {BR} (X \rightarrow  \ell\ell)$}}

\newcommand{\lw}[1]{\smash{\lower2.0ex\hbox{#1}}}
\newcommand{\lsqbr}   {{${\lambda^{\prime}}^{2}\cdot$Br}} 
\newcommand{\sneutrino} {{$\widetilde\nu$}}
\newcommand{\kmp}   {{$k/M_{\mathrm{PL}}$}}

\begin{document}

\renewcommand{\baselinestretch}{1.0}

\title{Search for New High Mass Particles Decaying to Lepton Pairs in
\boldmath{$p\bar{p}$} Collisions \\ at \boldmath{$\sqrt{s}=1.96$} TeV }

\affiliation{Institute of Physics, Academia Sinica, Taipei, Taiwan 11529, Republic of China} 
\affiliation{Argonne National Laboratory, Argonne, Illinois 60439} 
\affiliation{Institut de Fisica d'Altes Energies, Universitat Autonoma de Barcelona, E-08193, Bellaterra (Barcelona), Spain} 
\affiliation{Baylor University, Waco, Texas  76798} 
\affiliation{Istituto Nazionale di Fisica Nucleare, University of Bologna, I-40127 Bologna, Italy} 
\affiliation{Brandeis University, Waltham, Massachusetts 02254} 
\affiliation{University of California, Davis, Davis, California  95616} 
\affiliation{University of California, Los Angeles, Los Angeles, California  90024} 
\affiliation{University of California, San Diego, La Jolla, California  92093} 
\affiliation{University of California, Santa Barbara, Santa Barbara, California 93106} 
\affiliation{Instituto de Fisica de Cantabria, CSIC-University of Cantabria, 39005 Santander, Spain} 
\affiliation{Carnegie Mellon University, Pittsburgh, PA  15213} 
\affiliation{Enrico Fermi Institute, University of Chicago, Chicago, Illinois 60637} 
\affiliation{Joint Institute for Nuclear Research, RU-141980 Dubna, Russia} 
\affiliation{Duke University, Durham, North Carolina  27708} 
\affiliation{Fermi National Accelerator Laboratory, Batavia, Illinois 60510} 
\affiliation{University of Florida, Gainesville, Florida  32611} 
\affiliation{Laboratori Nazionali di Frascati, Istituto Nazionale di Fisica Nucleare, I-00044 Frascati, Italy} 
\affiliation{University of Geneva, CH-1211 Geneva 4, Switzerland} 
\affiliation{Glasgow University, Glasgow G12 8QQ, United Kingdom} 
\affiliation{Harvard University, Cambridge, Massachusetts 02138} 
\affiliation{Division of High Energy Physics, Department of Physics, University of Helsinki and Helsinki Institute of Physics, FIN-00014, Helsinki, Finland} 
\affiliation{University of Illinois, Urbana, Illinois 61801} 
\affiliation{The Johns Hopkins University, Baltimore, Maryland 21218} 
\affiliation{Institut f\"{u}r Experimentelle Kernphysik, Universit\"{a}t Karlsruhe, 76128 Karlsruhe, Germany} 
\affiliation{High Energy Accelerator Research Organization (KEK), Tsukuba, Ibaraki 305, Japan} 
\affiliation{Center for High Energy Physics: Kyungpook National University, Taegu 702-701; Seoul National University, Seoul 151-742; and SungKyunKwan University, Suwon 440-746; Korea} 
\affiliation{Ernest Orlando Lawrence Berkeley National Laboratory, Berkeley, California 94720} 
\affiliation{University of Liverpool, Liverpool L69 7ZE, United Kingdom} 
\affiliation{University College London, London WC1E 6BT, United Kingdom} 
\affiliation{Massachusetts Institute of Technology, Cambridge, Massachusetts  02139} 
\affiliation{Institute of Particle Physics: McGill University, Montr\'{e}al, Canada H3A~2T8; and University of Toronto, Toronto, Canada M5S~1A7} 
\affiliation{University of Michigan, Ann Arbor, Michigan 48109} 
\affiliation{Michigan State University, East Lansing, Michigan  48824} 
\affiliation{Institution for Theoretical and Experimental Physics, ITEP, Moscow 117259, Russia} 
\affiliation{University of New Mexico, Albuquerque, New Mexico 87131} 
\affiliation{Northwestern University, Evanston, Illinois  60208} 
\affiliation{The Ohio State University, Columbus, Ohio  43210} 
\affiliation{Okayama University, Okayama 700-8530, Japan} 
\affiliation{Osaka City University, Osaka 588, Japan} 
\affiliation{University of Oxford, Oxford OX1 3RH, United Kingdom} 
\affiliation{University of Padova, Istituto Nazionale di Fisica Nucleare, Sezione di Padova-Trento, I-35131 Padova, Italy} 
\affiliation{University of Pennsylvania, Philadelphia, Pennsylvania 19104} 
\affiliation{Istituto Nazionale di Fisica Nucleare Pisa, Universities of Pisa, Siena and Scuola Normale Superiore, I-56127 Pisa, Italy} 
\affiliation{University of Pittsburgh, Pittsburgh, Pennsylvania 15260} 
\affiliation{Purdue University, West Lafayette, Indiana 47907} 
\affiliation{University of Rochester, Rochester, New York 14627} 
\affiliation{The Rockefeller University, New York, New York 10021} 
\affiliation{Istituto Nazionale di Fisica Nucleare, Sezione di Roma 1, University di Roma ``La Sapienza," I-00185 Roma, Italy} 
\affiliation{Rutgers University, Piscataway, New Jersey 08855} 
\affiliation{Texas A\&M University, College Station, Texas 77843} 
\affiliation{Istituto Nazionale di Fisica Nucleare, University of Trieste/\ Udine, Italy} 
\affiliation{University of Tsukuba, Tsukuba, Ibaraki 305, Japan} 
\affiliation{Tufts University, Medford, Massachusetts 02155} 
\affiliation{Waseda University, Tokyo 169, Japan} 
\affiliation{Wayne State University, Detroit, Michigan  48201} 
\affiliation{University of Wisconsin, Madison, Wisconsin 53706} 
\affiliation{Yale University, New Haven, Connecticut 06520} 
\author{A.~Abulencia}
\affiliation{University of Illinois, Urbana, Illinois 61801}
\author{D.~Acosta}
\affiliation{University of Florida, Gainesville, Florida  32611}
\author{J.~Adelman}
\affiliation{Enrico Fermi Institute, University of Chicago, Chicago, Illinois 60637}
\author{T.~Affolder}
\affiliation{University of California, Santa Barbara, Santa Barbara, California 93106}
\author{T.~Akimoto}
\affiliation{University of Tsukuba, Tsukuba, Ibaraki 305, Japan}
\author{M.G.~Albrow}
\affiliation{Fermi National Accelerator Laboratory, Batavia, Illinois 60510}
\author{D.~Ambrose}
\affiliation{Fermi National Accelerator Laboratory, Batavia, Illinois 60510}
\author{S.~Amerio}
\affiliation{University of Padova, Istituto Nazionale di Fisica Nucleare, Sezione di Padova-Trento, I-35131 Padova, Italy}
\author{D.~Amidei}
\affiliation{University of Michigan, Ann Arbor, Michigan 48109}
\author{A.~Anastassov}
\affiliation{Rutgers University, Piscataway, New Jersey 08855}
% ...
\collaboration{CDF Collaboration}
\noaffiliation

\date{\today}

\begin{abstract}
A search for new particles ($X$) that decay to electron or muon pairs has been 
performed
using approximately 200 pb$^{-1}$ of~\mbox{$p\bar{p}$}~collision
data at $\sqrt{s}=1.96$ TeV collected by the CDF II experiment 
at the Fermilab Tevatron. 
Limits on 
$\sigxll$ are presented
as a function of dilepton invariant mass~$\mll > 150~\rm{GeV}/c^2$,
for different spin hypotheses (0, 1, or 2).  
The limits are approximately 25 fb for $\mll>600$~GeV/$c^2$.
Lower mass bounds for $X$~from representative models beyond the Standard Model 
including heavy neutral gauge bosons are presented.

\end{abstract}

\pacs{13.85.Rm, 12.38.Qk, 13.85.Qk, 14.70.Pw, 12.60.Cn}
\maketitle
\renewcommand{\baselinestretch}{1.0}

A search for new particles ($X$) has been performed
in the dilepton ($ee$ and $\mu\mu$) decay channel
using~\mbox{$p\bar{p}$}~collision
data at $\sqrt{s}=1.96$ TeV collected by the upgraded Collider Detector at 
Fermilab (CDF II) at the Tevatron.
The observed dilepton invariant mass ($\mll$) distribution is
compared with that expected
from Standard Model (SM) processes for $\mll > 150$~GeV/$c^2$.
Many models beyond
the SM predict such particles with masses at or below the TeV scale~\cite{langacker2}.
Generic searches for spin-0, 1, and 2 particles are performed, taking
into account the dependence of the experimental acceptance on the 
spin-dependent angular distributions of the lepton pair.
While this approach provides sensitivity to broad classes of new  models, 
the spin-1 result addresses an issue of fundamental importance in particle 
physics: the possible existence of extra neutral gauge bosons expected in 
many models with a higher gauge  structure than that of the SM.
A generic SM-like (sequential) $\zp$ boson ($\zpsm$) is defined to
have the same coupling strengths to fermions
as those of the SM $Z^{0}$ boson and its mass bound provides a convenient 
reference indicating the experimental sensitivity.  
The previous best 
$\zpsm$ lower mass bounds from direct searches are 690~GeV/$c^2$
by the CDF collaboration~\cite{zpprl1997} 
and 670~GeV/$c^2$ by the D0 collaboration~\cite{d0zp} at the 95\% confidence 
level (CL)~\cite{95cl}.
Increased integrated luminosity and center-of-mass energy for Run II are 
expected to 
provide a significant improvement over these previous results.
Indirect limits on the mass of $\zp$ bosons have been set by
the LEP II experiments~\cite{LEP2}.
A more detailed discussion of the LEP results and the advantages of the 
Tevatron search can be found in Reference~\cite{cucd}. 
In addition to $\zpsm$, we consider $\zp$ bosons (spin-1) from the $E_{6}$ 
model ($Z_{\chi}$,
$Z_{\psi}$, 
$Z_{\eta}$, 
$Z_{I}$)~\cite{E6} 
and the Littlest Higgs model ($Z_{H}$)~\cite{littlehiggs}, 
Technicolor (TC) particles (spin-1)~\cite{technimc},
sneutrinos (\sneutrino) in
an R-parity violating supersymmetric (RPV SUSY) model
(spin-0)~\cite{sneutrinos}, 
and gravitons in the Randall-Sundrum (RS) warped extra dimension 
model (spin-2)~\cite{rsgravitons}.
Independent of specific models, the limits on 
$\xlims \equiv \sigxll$
presented here can be used to set lower bounds on the mass of $X$ ($\mX$) in many
classes of models with a narrow width resonance.
Using the spin-1 $\xlims$ limit result, bounds on the couplings in more generalized
$\zp$ models~\cite{cucd} have been derived and are presented.


The CDF II detector is 
a forward-backward and azimuthally symmetric detector
with a tracking system immersed in a 1.4~T solenoidal magnetic field, 
calorimetry
for measuring the energies of particles, and detectors to identify 
deeply-penetrating muons~\cite{detector}. 
The tracking system consists of an open-cell drift chamber, the Central Outer 
Tracker (COT), surrounding an eight layer silicon tracker.  The fiducial 
coverage of the COT is $|\eta|<1.0$ and
the silicon extends this coverage forward to $|\eta|<1.8$~\cite{CDFcoo}.
The tracking system is surrounded by 
electromagnetic (EM) and hadronic (HAD) calorimeters that are divided into 
a central calorimeter ($|\eta|<1.1$) and two forward, or ``plug'', calorimeters 
($1.2<|\eta|<3.6$).   
Drift chambers, located outside the hadronic calorimeters and also outside
an additional 60 cm of iron shield, detect muons having 
$|\eta|<1.0$.


Candidate events are selected from data collected during 2002 - 2003, 
corresponding to an integrated luminosity ranging from 173 to 200 pb$^{-1}$, 
depending upon 
the detector elements required for the analysis.
Dielectron events with a central candidate are 
collected using 
a single-electron trigger requiring a loosely-selected electron in the central
EM calorimeter (CEM) with 
$\etm>18$~GeV and a matching COT track with
$\ptm>9$~GeV/$c$. 
Dielectron events 
without a central candidate are collected using a trigger 
requiring two 
loosely-selected 
electron candidates in the plug EM calorimeter (PEM) 
with $\etm > 18$~GeV and no tracking requirement.  
Additional triggers with higher $\etm$ thresholds but looser 
electron-selection requirements 
are used to ensure full efficiency for high-mass events.  Together, 
these triggers are essentially 100\% efficient for the $ee$ decay mode
for $\mll > 150$~GeV/$c^2$.  
Dimuon candidate events are collected with single-muon
triggers which require 
a muon chamber track with a matching track measured by the 
COT with $\ptm>18$~GeV/$c$.
The overall trigger efficiency for the $\mu\mu$ decay mode is above 90\%.

The dilepton event selection requires at least two electron or two
muon candidates with no charge requirement. Both electron and muon
candidates are required to be isolated with a cut on the energy found
within a cone of angular radius 
$R=\sqrt{{(\delta\phi)}^2 + {(\delta\eta)}^2}=0.4$ around the lepton
candidate. 
Electron candidates require an EM cluster with $\etm > 25$
GeV and longitudinal and transverse shower profiles consistent with
electrons~\cite{ref_WZsigma}. 
At least one of the two
electrons is required to have a matching track, except for events with
two central electrons, which both require matching tracks. The inclusion of
events with two forward electrons is possible due to a
calorimeter-seeded forward tracking algorithm~\cite{ref_Wasym}. Events with
a significant amount of \met are rejected
to remove W+jets and others backgrounds with unreconstructed
particles.
All muon candidates are 
required to have a
COT track with $\ptm>20$~GeV/$c$ and calorimeter energy deposition 
consistent with a minimum-ionizing particle signal, where at least
one candidate must also have a matching 
track in the muon chambers. 
To reject cosmic-ray events, muon candidates are required to 
have COT hit-timing consistent with outward-moving particles~\cite{cosmiccut}.

\begin{figure}
\vspace*{-1.0cm}
\includegraphics[scale=.6]{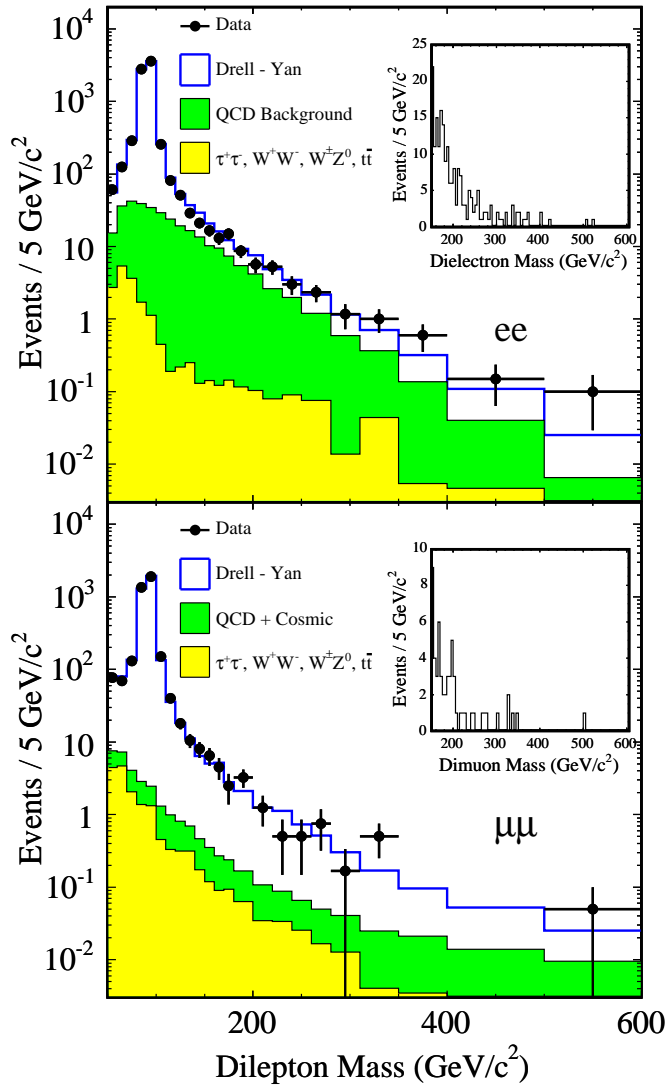}
\caption{\label{fig:Dyeemumufig}
The $ee$ (top) and $\mu\mu$ (bottom) invariant mass distributions of the 
observed data (points)
with the background prediction (solid line). The background is corrected for 
acceptance and efficiency.  The insets show the data with a fixed bin width
of 5~GeV/$c^2$ for $\mll>150$~GeV/$c^2$.
}
\end{figure}
%
The selected data contains 14,799 $ee$ and 7,775 $\mu\mu$ candidate events 
with the 
dilepton invariant mass distributions shown in Fig.~\ref{fig:Dyeemumufig}.  
These samples are dominated by events in the $Z^0$ peak. In this region 
the dielectron sample has a larger acceptance; however, in the high-mass 
search region the two channels have similar sensitivity.
The lepton identification efficiencies are estimated using
a purified sample of dilepton events
from $Z^{0}$ decays~\cite{zpprl1997}.
Since leptons from the decay of high-mass objects 
typically have higher
$\ptm$ than this sample, the lepton identification efficiency
is studied as a function of $\ptm$, and
the selection criteria are chosen to 
ensure high efficiencies throughout the relevant $\ptm$ 
range~\cite{koji,muge}. 
The geometric and kinematic acceptance as a function of
resonance mass is estimated using Monte Carlo (MC) samples: the PYTHIA event 
generator~\cite{pythia} with CTEQ5L parton distribution 
functions (PDF)~\cite{CTEQ5L} and the CDF~II detector simulation are 
used except as noted.
Signal samples for the heavy Higgs (spin-0), 
\zpsm\ (spin-1)
and RS Graviton (spin-2) are generated to model each spin hypothesis.
The product of acceptance and selection efficiency is approximately 50\%
for $\mX > 400$~GeV/$c^2$ for $ee$ and $\mu\mu$ for all spins.
\begin{table}
\caption{\label{Dytable}Integrated number of events above a given $\mll$
for the observed data and estimated background.}
\begin{ruledtabular}
\begin{tabular}{c cc cc}
$\mll$            &  \multicolumn{2}{c}{$ee$ } &
                     \multicolumn{2}{c}{$\mu\mu$}\\
(GeV/${c}^{2}$)   & Observed & Expected    & Observed    & Expected \\
\hline
$>$~150    & 205  & 212.9 $\pm$ 99.3 & 58 &  55.3 $\pm$ 2.5\\
$>$~200    &  84  &  78.2 $\pm$ 33.4 & 18 &  20.9 $\pm$ 1.0\\
$>$~300    &  22  &  13.6 $\pm$  4.4 &  6 &   5.2 $\pm$ 0.3\\    
$>$~400    &   5  &   2.9 $\pm$  0.7 &  1 &   2.3 $\pm$ 0.2\\
$>$~500    &   2  &   0.8 $\pm$  0.1 &  1 &   1.2 $\pm$ 0.1\\
\end{tabular}
\end{ruledtabular}
\end{table}
%
%
\begin{figure*}
\includegraphics[scale=.85]{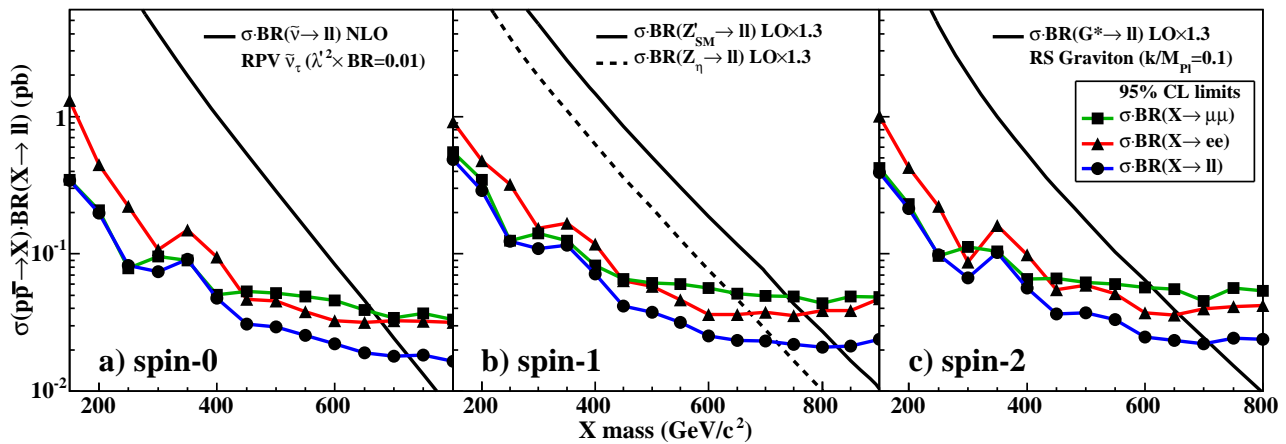}
\caption{\label{fig:limits}
The \xlims~ limits from $ee$, $\mu\mu$ and the combined channels
as a function of $\mX$ for spin-0 (a),
spin-1 (b), and spin-2 (c).  For the combined channel, 
$BR (X\to ee) = BR (X\to \mu\mu)~(\equiv {BR(X\to \ell\ell)})$
is assumed.
Also shown are theoretical cross-section predictions of 
representative models. 
}
\end{figure*}
%
\begin{table*}
\caption{\label{limit_values}
95\% CL upper limits on 
$\sigxll$ (in fb) 
for a given $\mX$ (in GeV/$c^2$).
Spin-1 limits are computed to 900 GeV/$c^2$ to accommodate \zp  models  
with large predicted cross-sections. 
}
\begin{ruledtabular}
\begin{tabular}{ccccccccccccccccc}
Spin$\backslash\mX$  & 150 & 200 & 250 & 300 & 350 & 400 & 450 & 500 & 550 & 600 & 650 & 700 & 750 & 800 & 850 & 900 \\  \hline
Spin-0      & 340 & 200 &83 &74 &91 &48 &31 &29 &26 &22 &19 &18 &18 &17 & ---  & ---  \\
Spin-1      & 490 & 290 & 120 & 110 & 120 &72 &42 &38 &32 &25 &24 &23 &22 &21 &21 & 24 \\ 
Spin-2      & 390 & 210 &98 &67 & 100 &56 &37 &37 &33 &25 &24 &22 &24 &24
 & ---  & ---  \\
\end{tabular}
\end{ruledtabular}
\end{table*}
%
%

The primary and irreducible SM background results from Drell-Yan production of
$ee$ and $\mu\mu$ pairs. It is estimated using MC simulation normalized to 
fit to the data in the $Z^0$ peak, 
after the other background contributions have been subtracted.
This reduces the effect of the luminosity uncertainty on the
background estimate.
The other contributions
such as $t\bar{t}$ (generated with HERWIG~\cite{herwig}), $\tau^+\tau^-$, 
$W^+W^-$, and $W^{\pm}Z^0$ are estimated using MC simulation.
Some accepted $ee$ events come from non-dielectron sources, predominantly 
misidentified QCD dijet events. 
This background is estimated by extrapolating from events
where the leptons are not isolated.
The QCD background in the $\mu\mu$ channel is estimated using same-sign events 
that pass the selection criteria and is found to be small.  
The cosmic ray background in the $\mu\mu$ channel is estimated
by applying the signal selection criteria to a sample of cosmic ray data
collected by the CDF II detector and is non-negligible at high 
mass ($\mll > 400$~GeV/$c^2$).
Fig.~\ref{fig:Dyeemumufig} compares the estimated background distributions to the
$ee$ and $\mu\mu$ data.  Table~\ref{Dytable} shows the integrated  number of events 
observed and expected above a given $\mll$.


Systematic uncertainties on the acceptance, efficiency and luminosity result in a 
relative 
uncertainty on the scale of \xlims~of approximately 10\%.  The largest contributions are 
from the uncertainties on luminosity, energy/momentum scales and resolutions, and the 
choice 
of PDF as estimated by comparison of different PDF parameterizations.  
Background uncertainty in the $ee$ channel
ranging from 40-80\% due to misidentified jets
results in absolute uncertainties on values 
of \xlims~that are large for $\mll<350$~GeV/$c^2$ but negligible at the higher 
mass region. Background uncertainties in the $\mu\mu$ channel are $\approx30$\%
and $\approx20$\% due to fake muons and cosmic-rays respectively. 
The relative uncertainty with respect to the scale of \xlims~ 
on the electroweak backgrounds is $\approx5$\%~in the both channels.

\begin{figure}
\includegraphics[scale=.5]{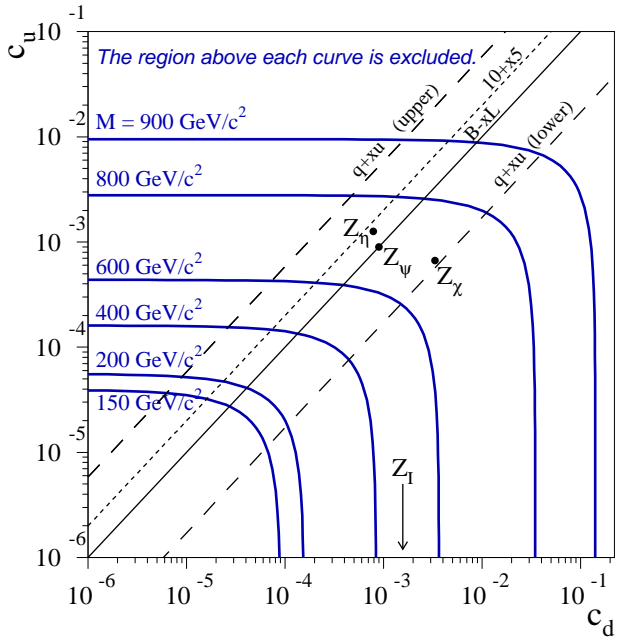}
\caption{\label{fig:cucd_contours}
Limit contours in the ($c_d$, $c_u$) plane~\cite{cucd} for a given~\zp~mass 
derived from the spin-1 $\xlims$ limit.  
The solid and dotted diagonal lines show all 
possible models for the ${U\rm{(1)}}_{B-xL}$ and
${U\rm{(1)}}_{10+x\bar{5}}$ groups respectively. The two dashed lines show the range
between which the values for the ${U\rm{(1)}}_{q+xu}$ group must fall.
The values for the ${U\rm{(1)}}_{d-xu}$ group may fall anywhere on the plane.  
The parameters of the $E_6$-model \zp~bosons are indicated.
}
\end{figure}

Since no significant excess of events is observed, limits on
\xlims~are extracted using a Bayesian, binned 
likelihood method.
For combined dilepton results assuming $BR (X\to ee) = BR (X\to \mu\mu)$, a 
joint likelihood is formed
from the product of the individual-channel likelihoods accounting for the 
correlations among 
systematic uncertainties.  
When the nuisance parameters are integrated out,
uncertainties on PDF, 
luminosity and common selection efficiencies are taken as 100\% correlated 
among the different components of the acceptance.
This joint likelihood is converted to a posterior density in 
the signal cross section and numerically integrated to
obtain the 95\% CL limits on $\xlims$.
Fig.~\ref{fig:limits} and Table~\ref{limit_values} show the $\xlims$ limits
as a function of $\mX$ with spins-0, 1, and 2. At high mass 
($\mX > 600$~GeV/$c^2$)
the limits are approximately 25 fb for all spins (but best for spin-0) 
and are consistent with expected limits 
in the absence of signal. The corresponding CDF Run I limit was 
40 fb~\cite{zpprl1997}. 
The sensitivity of these searches is enhanced compared to the Run I 
searches
by the addition of the plug-plug dielectrons (10\% relative gain in $ee$ acceptance), 
an increase in muon trigger coverage 
and the use of muons without muon-chamber tracks (50\% relative gain 
in $\mu\mu$ acceptance).
Fig.~\ref{fig:limits} also shows the predictions from representative models
with higher order corrections~\cite{kfactor}.
The particle 
$X$ is assumed to decay only to the known fermions in the mass range examined. 
From the spin-0 $\lims$ shown in Fig.~\ref{fig:limits}(a),
the lower mass bounds of
680, 620, and 460~GeV/$c^2$ from $ee$ channel and
665, 590, and 450~GeV/$c^2$ from $\mu\mu$ channel
are obtained for \sneutrino~
for the coupling strength squared times branching fraction 
(\lsqbr)= 0.01, 0.005, and 0.001 respectively.
For spin-1 (Fig.~\ref{fig:limits}(b))
the following mass bounds are
obtained from the combined channel:
825, 690, 675, 720 and 615~GeV/$c^2$ for $\zpsm$, 
$Z_{\chi}$,
$Z_{\psi}$, 
$Z_{\eta}$ and
$Z_{I}$ respectively
and
885, 860 ,805 and 725~GeV/$c^2$ for
$Z_{H}$ with the mixing parameter cot$\theta_{H}$ = 1.0, 0.9, 0.7 and
0.5 respectively.
%
%
%
%
Similarly, the lower mass limits of 280~GeV/$c^2$ 
(270~GeV/$c^2$) are set for $\rho_{TC}$ and  $\omega_{TC}$ in the 
TC model~\cite{technimc}  
with corresponding values of Technicolor-scale mass parameters 
$M_{V} = M_{A}$ of 500~GeV/$c^2$ (400~GeV/$c^2$).
From the spin-2 $\lims$ shown in Fig.~\ref{fig:limits}(c), 
the lower mass bounds of
710, 510, and 170~GeV/$c^2$ are obtained for the first excited
state of the RS graviton for dimensionless coupling parameter (\kmp) 0.1, 0.05,
and 0.01 respectively, where $k$ is the relative strength of the warped
dimension's curvature scale and $M_{\mathrm{PL}}$ is the effective Planck scale.
A method of factorizing the couplings, charges and 1/$s$ dependence of \zp~cross sections
from kinematic factors that depend upon PDF parameterizations allows more general 
constraints on
possible \zp~models~\cite{cucd}.  In this formalism, a generic \zp~is described by two 
parameters, 
$c_d$ and $c_u$, that define the coupling of down and up-type quarks to the resonance.
Fig.~\ref{fig:cucd_contours} shows the bounds set by the spin-1 limits in the 
($c_d$, $c_u$) 
plane along with the parameters describing the four $E_6$-model \zp~bosons.

We thank D.~Choudhury, A.~Daleo, H.~Logan, 
and S.~Mrenna for their useful contributions.
We thank the Fermilab staff and the technical staffs of the participating
institutions for their vital contributions. This work was supported by
the U.S. Department of Energy and National Science Foundation; the
Italian Istituto Nazionale di Fisica Nucleare; the Ministry of Education,
Culture, Sports, Science and Technology of Japan; the Natural Sciences
and Engineering Research Council of Canada; the National Science Council
of the Republic of China; the Swiss National Science Foundation; the A.P.
Sloan Foundation; the Bundesministerium f\"ur Bildung und Forschung,
Germany; the Korean Science and Engineering Foundation and the Korean
Research Foundation; the Particle Physics and Astronomy Research Council
and the Royal Society, UK; the Russian Foundation for Basic Research;
the Comisi\'on Interministerial de Ciencia y Tecnolog\'{\i}a, Spain; in
part by the European Community's Human Potential Programme under contract
HPRN-CT-2002-00292; and the Academy of Finland.


\renewcommand{\baselinestretch}{1.0}
\onecolumngrid


\begin{thebibliography}{0}
\bibitem{95cl}All the limits presented in this paper are at the 95\% CL.
\bibitem{CDFcoo} 
CDF uses a cylindrical coordinate system in which 
$\phi$ is the azimuthal angle, and
$+ z$ points in the direction of the proton beam and is zero at the
center of the detector.
The pseudorapidity $\eta~\equiv~-$ln(tan($\theta$/2)), where $\theta$ is the polar angle
relative to the $z$ axis. Calorimeter energy (track momentum) measured
transverse to the beam is denoted as $\etm$ ($\ptm$), and the total
calorimeteric transverse energy imbalance is denoted as $\met$.
\bibitem{kfactor}
A constant K-factor of 1.3 is used to be consistent with the previous analyses
making comparison of the \zp mass limits easier. The NLO calculation is used for
the RPV \sneutrino~case.
The dependence on the higher order corrections for the $\xlims$ limits is negligible.
\end{thebibliography}
\end{document}